# Generating strong cylindrical vector pulses via stimulated Brillouin amplification

Zhi-Han Zhu,[1,a),b)] Peng Chen,[2,a)] Li-Wen Sheng,[3] Yu-Lei Wang,[3] Wei Hu,[2,c)] Yan-Qing Lu,[2] and Wei Gao[1]

[1]*Institute of Photonics and Optical Fiber Technology, Harbin University of Science and Technology, Harbin 150080, China.*

[2]*National Laboratory of Solid State Microstructures, Collaborative Innovation Center of Advanced Microstructures and College of Engineering and Applied Sciences, Nanjing University, Nanjing 210093, China.*

[3]*National Key Laboratory of Science and Technology on Tunable Laser, Harbin Institute of Technology, Harbin 150001, China.*

Light with transverse polarization structure, such as radial and azimuthal polarization, enables and revives lots of applications based on light-matter interaction due to their unique focal properties. To date, studies referring to this topic mainly concentrate in weak-light domain, yet it should have gained more attention in strong-light domain. A main factor contributing to the current situation is that the generation devices cannot afford high power. Here, in this proof-of-principle work, we demonstrate the generation of strong single-longitudinal-mode (SLM) cylindrical vector (CV) short pulses via stimulated Brillouin amplification, specifically, the energy is transferred from a 700 ps pump to a 300 ps Stokes pulse via parametrically generated coherent phonons. After amplification, a 100 mJ-level SLM CV pulse light with 300 ps duration is obtained. Meanwhile, the phase and polarization structures are high fidelity maintained. This result provides a practicable way to generate strong CV light, and by extending this mechanism into multi-pump beam combing system, even an ultra-high intensity CV light can be expected.

Science of structured light—the generation and application of light fields with custom transverse intensity, phase and polarization profiles—has gained enormous progress and gradually become the hottest topic of optical community over the past decade.[1,2] In this new subject of modem optics, orbital angular momentum (OAM)[3] plays a crucial role, which is associated with light's twisted phase profile. Particularly, the infinite dimensional space spanned by this degree-of-freedom (DOF) has directly given rise to a promising research direction of the topic—informatics, including both classical and quantum domains.[4] In addition to intensity and phase structures, the polarization of light can also exhibit a non-uniform transverse distribution once the OAM and spin angular momentum (SAM) DOFs are entangled.[5,6] Among them, radially and azimuthally polarized light, two specific members of the so-called cylindrical vector (CV) beam, have received more attention, because their focal

---

regions manifest a strong longitudinal electric field and magnetic field, respectively.[7] This striking focal property, such as allowing smaller focal spot, inspires another fascinating research direction of the structured photonics—light-matter interaction, ranging from measurement, imaging, microscopy to optical manipulation at the nanoscale.[2] In this domain, pulsed light with single longitudinal mode (SLM) is welcome for many applications, especially for high-power applications that expect the ability of shaping focal property more.[8] However, it is still a challenge for the key technique of CV light generation devices in current stage, such as q-plates and metasurfaces based on spin-orbital coupling,[9,10] whose damage thresholds are low and particularly for SLM light (i.e., all photons in one quantum state means a larger complex amplitude).

An alternative method to generating high-power CV pulses is employing light amplification techniques, such as laser amplification (LA), optical parametric amplification (OPA), and energy transfer via stimulated Raman or Brillouin scattering (SRS and SBS).[11] For SLM pulses with sub-nanosecond duration, the efficiency of LA is lower, while OPA needs specific pump source and larger aperture, making whole system complicated. In contrast, amplifying light pulses via SRS and SBS in gases, liquids and plasmas have many benefits. For instance, liquids and gases media provide much higher damage threshold, and plasmas possess almost a damage-free feature that has been used in cross-beam energy transfer scheme at the National Ignition Facility.[12] So far, amplifications of OAM light via SRS and SBS have been proved.[13,14] In this work, we propose and demonstrate the generation of strong SLM CV short pluses via stimulated Brillouin amplification (SBA) in fluorocarbon liquid.

The high-energy SLM picosecond source is provided by a SBS compression Nd:YAG laser system. The core of this system is a diode pumped Q-switched nanosecond pulses generator, and SLM selection property is achieved by employing geometry layout of the output mirror and an optical flat, a more cost-efficient and simple way compared to Fabry-Perrot etalon. Then, the pulse is directed to double-pass laser amplifier with a SBS based phase-conjugation mirror. After the amplification and SBS compression, the final output are 250 mJ SLM pulses with 700 ps duration and 3 mm diameter. Partial of this output is directed and focused into a SBS cell to generate 300 ps Stokes frequency shift seed pulses required for the next step. Here, we focus on the radial CV light that can be expressed as a superposition of two Laguerre-Gauss (LG) modes with opposite topological charges and circular polarizations, i.e., $(|L,-LG_{01}\rangle+|R,+LG_{01}\rangle)/\sqrt{2}$, or two Hermite-Gauss (HG) modes in orthogonally linear polarizations, i.e., $(|H,HG_{10}\rangle+|V,HG_{01}\rangle)/\sqrt{2}$.[6] To amplify this CV light via SBA, evidently, a pair of linearly (H+V) or circularly (L+R) polarized pumps with same energy and phase-matching condition is necessary.

Figure 1 shows the schematic illustration of this proof-of-principle experiment. A non-collinear SBA layout is adopted and the cross angle $\theta$ between the two pumps and the seed is set as $\theta = \pm6°$, providing enough interaction length and acceptable phase-matching condition.[13,15] The CV seeds are converted from a linearly polarized 300 ps SLM pulses by passing through a q-plate with q = 0.5, and the highest energy of the seed pulse is set as 20 mJ, an affordable power for liquid-crystal based spin-orbital devices at 1064 nm. Then, except for the difference in polarization, two identical pumps with 700 ps duration interact with the seed pulse in a 10 cm-length coupling cell contained FC-72, which is widely used in SBS compression laser system. Two quarter-wave plates inserted in pump paths are used for converting pump light into required polarization combination (H+V or L+R). After the amplification, the high-energy output pulses are analyzed by polarization tomography consisted of two wave plates and a polarized beam splitter (PBS). The transverse profiles and energies of output pulses before and after the polarization tomography are recorded by a CCD and an energy meter, respectively.

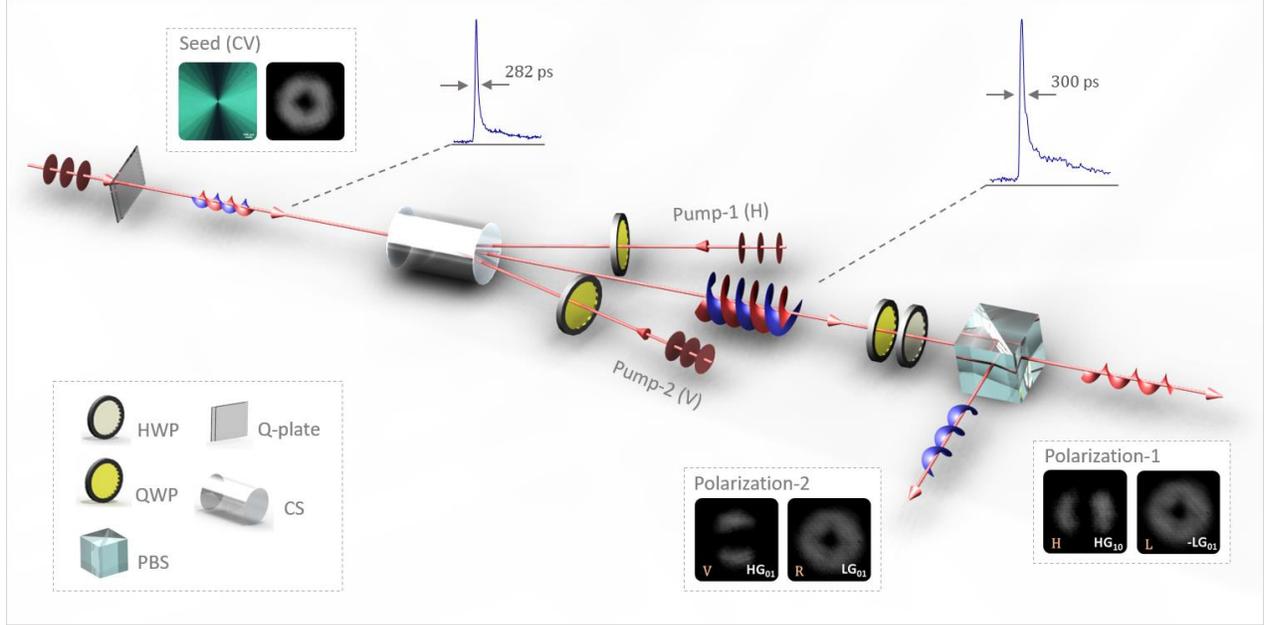

FIG. 1. Schematic illustration of the experimental setup. Key components include q-plate, half-wave plate (HWP), quarter-wave plate (QWP), polarized beam splitter (PBS), and coupling cell (CS). A 300 ps-level CV SLM pulse light (Stokes shifted) interacts with pump-1 and pump-2 in the coupling cell simultaneously. The polarizations of the two pumps are perpendicular to each other, and the cross angles between them are 6°. Two wave plates and a PBS shown at lower right-hand constitute a polarization tomography module that analyzes energies and transverse profiles of the output signals.

The configuration of SBA for radial CV light shown in Fig. 1 can be regarded as two mutually independent photon-phonon coupling processes, or rather, two pumps respectively amplify corresponding polarization (half) of the seed. One may point out that amplifying CV light via only one photon-phonon coupling could be achieved by using CV pump, nevertheless, this assumption neglects a fact that generation high power CV pumps was actually a bottleneck. The phase-matching condition for a non-collinear SBA used here can be expressed as $\mathbf{q}(\Omega') = \mathbf{k}(\omega_p) - \mathbf{k}(\omega_s)$, where $\mathbf{k}(\omega)$ and $\mathbf{q}(\Omega')$ are the optical and acoustic dispersion relations, respectively, and $\Omega' = \cos(\theta/2)(\omega_p - \omega_s)$ is phonons' frequency generated under this configuration. For such a small cross angle ($\theta = 6°$), $\Omega' \approx \Omega = \omega_p - \omega_s$ (1.1 GHz for FC-72) and consider that the line-widths of the utilized sub-ns pulses are large, the decrease of the coupling efficiency induced by the phase-mismatching is negligible. Besides, compared to all-light based OPA, a major defect of parametric amplification via light-matter interaction is the noise arising from the interactions between optical fields and non-coherent collective excitations. While the non-collinear SBA scheme directly avoids this thorny problem, owning to a cross angle existed between the propagation directions of the output signals and the noise. In our experiment, the seed energy is set as 5 mJ, 10 mJ, and 20 mJ, respectively, and the energy of the two pumps are set from 10 mJ to100 mJ synchronously with a step of 10 mJ. Figure 2(a) shows the dependence of output energy upon the input one-way pump energy. The results indicate that total output energy increases linearly with the increase of input pump energy and a higher seed can extract more energy from pumps in same condition. The embedded inset in the top left of Fig. 2(a) shows the average final output energy when the input pump energy is 100 mJ × 2. We can see that 100 mJ-level SLM CV pulses with 300 ps duration are obtained when the interaction is seeded by 20 mJ input, corresponding to an approximate 4.6 GW/cm$^2$ power density. In addition, it can be noted that energy in polarization-1 is slightly smaller than that in polarization-2. This phenomenon ascribes to an energy difference between the two pump, and a finer design upon demand can solve this problem.

Figure 2(b) shows the energy transfer efficiency ($\eta$) from pump to Stokes for different pump energies. As expected, in the same input pump energy, a higher energy ratio of the seed to pump ($E_s/E_p$) can obtain a higher $\eta$. Moreover, $\eta$ increases with the increasing of pump energy for the same seed condition and $\eta$ becomes saturated as pump reaching 70 mJ. Although $\eta$ is only achieved up to 40% level under this proof-of-principle configuration, it can be improved rapidly when introducing another group of pumps in the next amplification stage, i.e., so-called SBS beam combining, due to the increase of $E_s/E_p$. Usually, $\eta$ can exceed 80% when $E_s/E_p$ is greater than 4 in fluorocarbon based SBA system in high energy range,[16] providing a feasible way to generate an ultra-high intensity CV pulses.

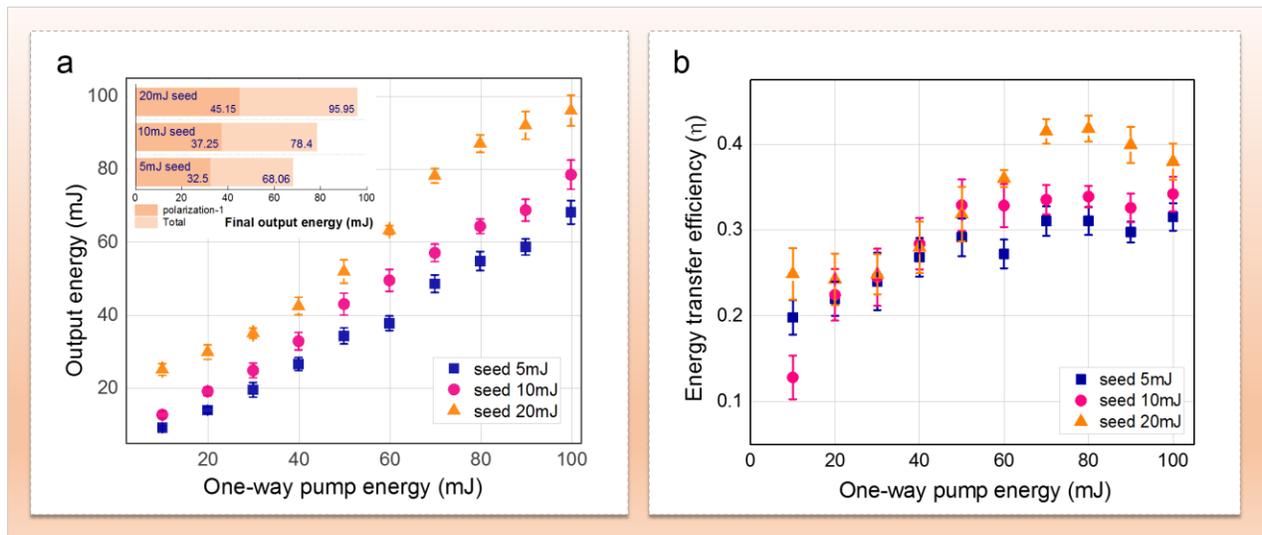

FIG. 2. (a) The output energy of amplified seed pulses versus input one-way pump energy. The top left inset shows the final output energy when input pump energy is 100 mJ, including total energy and energy in polarization-1. (b) The energy transfer efficiency ($\eta$) from pump to Stokes versus the pump energy. The blue, pink and orange points shown in (a) and (b) correspond to 5 mJ, 10 mJ and 20 mJ input seed energy, respectively.

Data shown in Fig. 2 involve two cases, i.e., H+V and L+R pumps. As expected, there is no difference reflected in output energy between adopting orthogonally linear polarizations and opposite circular polarizations. This situation is also proved by measuring transverse profiles, with the polarization tomography of the final output pulses with 100 mJ pumped shown in Fig. 3. It can be seen that the output pulses seem to be a HG or LG mode under one-way pumped, as shown in Figs. 3(a), (b), (d) and (e). This is because that after the amplification, only pump-1 input, the original seed field evolves into $\alpha|L,-LG_{01}\rangle+\beta|R,+LG_{01}\rangle$ or $\alpha|H,HG_{10}\rangle+\beta|V,HG_{01}\rangle$ where $|\alpha|^2+|\beta|^2=1$ and $\alpha:\beta\approx 5:1$. Thus, comparing the amplified component with unamplified component, the latter becomes too weak under the same attenuation level. Once the two pumps input simultaneously, as shown in Figs. 3(c) and (f), the output pulses manifest a typical characteristic of radial CV light indicated by corresponding polarization tomography.

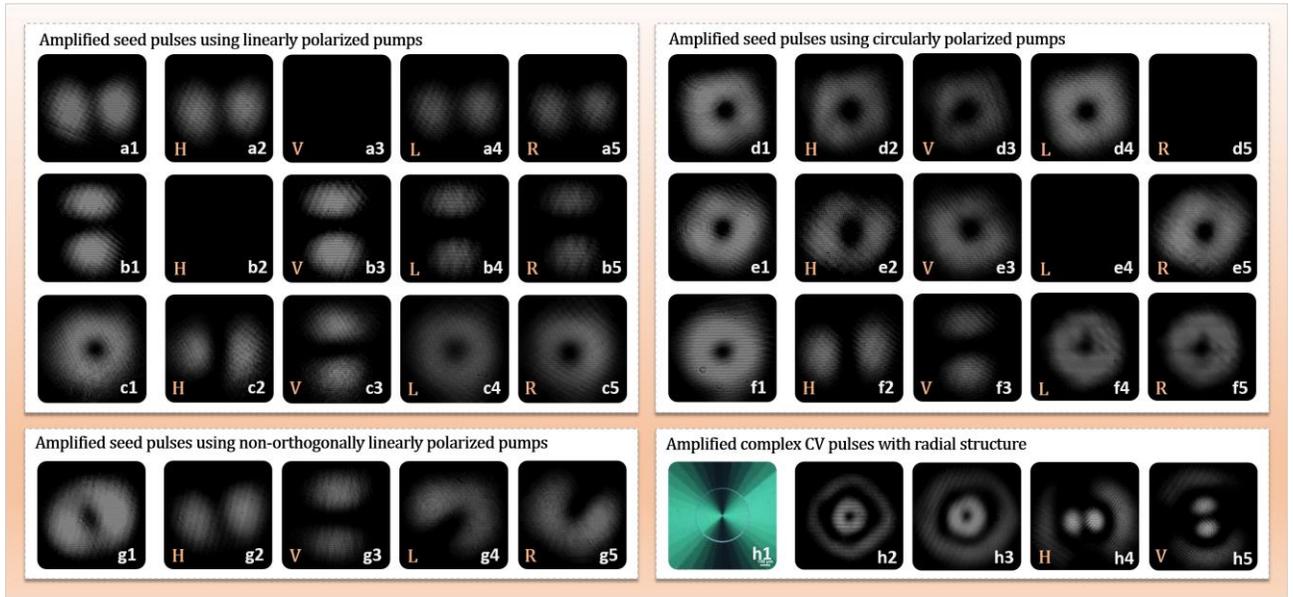

FIG. 3. Transverse profiles and polarization tomography of the final output pulses. (a)-(c) The seed is amplified by linearly polarized pumps, (a1)-(a5) only input H polarized pump; (b1)-(b5) only input V polarized pump; (c1)-(c5) both H and V pumps are input. (d)-(f) The seed is amplified by circularly polarized pumps, (d1)-(d5) only input L polarized pump; (e1)-(e5) only input R polarized pump; (f1)-(f5) both L and R pumps are input. (g) Transverse profiles and polarization tomography of the final output pulses amplified by two non-orthogonal linearly polarized pumps. (h1) Micrograph of a q-plate with q = 0.5 and p = 1, (h2) Transverse profile of the input complex CV pulses (h3)-(h5) Transverse profiles and polarization tomography of the final output complex CV pulses.

Notice that in order to achieve a high fidelity amplification, a key point is the orthogonality between the two pumps, or the output light field's structure in polarization and spatial DOF will distort. For instance, as shown in Fig. 3(g), after an amplification performed by two non-orthogonal linearly polarized pumps, the donut transverse profile become non-uniform and its polarization tomography in L and R projection even manifest LG modes with fractal topological charges. In addition to the standard radial and azimuthal pulses, SBA can also be applied for amplifying more complex CV light.[10,17] Figure 3(h) shows the polarization tomography of amplified CV pulses with a radial structure. By employing such high-energy SLM complex CV pulses, a more flexible focal property can be provided for the research on light-matter interaction.

In summary, we have demonstrated the generation of high-energy SLM CV short pulses via SBA, and a 100 mJ-level radial CV light with 300 ps pulse duration is achieved. After the amplification, measurements of polarization tomography in energy and transverse profile indicate that the polarization structures are high fidelity maintained. These results suggest that the SBA is a feasible way to generate high power CV light, and ultra-high intensity CV pulses could be generated by extending this configuration into multi-pumped beam combining system. What's more, this mechanism can also work in plasmas based SBA system, to generate ultra-high energy CV fs pulses.

This work is supported by the National Natural Science Foundation of China (Grant Nos. 61378003, 61490714, 61575093, 11574065), the Key Programs of the Natural Science Foundation of Heilongjiang Province of China (Grant No. ZD201415).